\documentclass[a4paper, 12pt,psfig]{iopart}
\usepackage{iopams}
\usepackage{epsf}
\usepackage{euscript}
\usepackage{amsthm}
\usepackage{setstack}
\usepackage{graphicx}
\usepackage{color}
\usepackage{bm}
\usepackage{savesym}

\makeatletter

\newcommand{\Rmnum}[1]{\expandafter\@slowromancap\romannumeral #1@}
\makeatother

\begin{document}
\title{Trace Formulae and Spectral Statistics for Discrete Laplacians on Regular Graphs ($\Rmnum{2}$)}
\author{Idan Oren$^{1}$ and Uzy Smilansky$^{1,2}$}
\today
\address{$^{1}$Department of Physics of Complex Systems,
Weizmann Institute of Science, Rehovot 76100, Israel.}
\address{$^{2}$School of Mathematics, Cardiff University, Cardiff,
Wales, UK}
 \ead{\mailto{idan.oren@weizmann.ac.il}\\
 \mailto{\ \ \ \ \ \ \ uzy.smilansky@weizmann.ac.il}}

\begin{abstract}
Following the derivation of the trace formulae in the first paper in
this series, we establish here a connection between the spectral
statistics of random regular graphs and the predictions of Random
Matrix Theory (RMT). This follows from the known Poisson
distribution of cycle counts in regular graphs, in the limit that
the cycle periods are kept constant and the number of vertices
increases indefinitely. The result is analogous to the so called
``diagonal approximation" in Quantum Chaos. We also show that by
assuming that the spectral correlations are given by RMT to all
orders, we can compute the leading deviations from the Poisson
distribution for cycle counts. We provide numerical evidence which
supports this conjecture.

\end{abstract}

\section{Introduction}
  \label{sec:intro}
The present paper is the second in this series, where we aim to
establish a rigorous connection between the spectral fluctuations in
the spectra of random regular graphs and the predictions of Random
Matrix Theory (RMT). In the first paper \cite{IdanAmitUzy} (to be
referred to as $\Rmnum{1}$) we provided the necessary definitions
and facts about graphs and we shall use the same notations here.
Suffice it to say that we deal with the ensemble $\mathcal{G}_{V,d}$
of $d$-regular graphs on $V$ vertices, and we study the spectrum
$\{\mu_j\}_{j=1}^{V-1}$ of the adjacency matrix $A$, from which we
excluded the trivial eigenvalue $\mu_V=d$. The spectral density is
defined as
\begin{equation}
\rho^{(A)}(\mu)  \equiv
\frac{1}{V-1}\sum_{j=1}^{V-1}\delta(\mu-\mu_j)\ .
 \label{eq:density}
\end{equation}
In what follows we shall be interested in the large $V$ limit, and
in most cases the replacement of $V-1$ by $V$ will be justified. We
shall do this consistently to simplify the notation.

 In $\Rmnum{1}$ we also defined the ensemble of
``magnetic" graphs, $\mathcal{G}^{M}_{V,d}$ where the adjacency
matrix of each member of the ensemble $\mathcal{G}_{V,d}$ is
decorated by phases
\begin{equation}
 M_{i,j}= A_{i,j} {\rm e}^{i\chi_{i,j}} \ \ ;\ \
\chi_{j,i}=-\chi_{i,j}
 \label{magneticA}
 \end{equation}
and the phases are independent random variables distributed
uniformly on the unit circle. Here the entire spectrum of $M$ is
considered and
\begin{equation}
\rho^{(M)}(\mu)  \equiv \frac{1}{V}\sum_{j=1}^{V}\delta(\mu-\mu_j)\
.
 \label{eq:densitym}
\end{equation}

In $\Rmnum{1}$ we prepared the tools needed for our purpose, namely
trace formulae. In the sequel, we shall summarize the absolutely
necessary information about trace formulae required to make the
present paper self contained.

\subsection {The trace formula - a short reminder}
Trace formulae express the spectral density of the adjacency matrix
as a sum of two contributions, $\rho(\mu) \equiv
\frac{1}{V}\sum_{j=1}^{V}\delta(\mu-\mu_j) = \left
\langle\rho(\mu)\right \rangle +\tilde{\rho}(\mu)$ where $\left
\langle\ \cdot\ \right \rangle$ stands for the ensemble average.
Here, For both ensembles,  $\left \langle\rho(\mu)\right \rangle$ is
the well known Kesten-McKay expression for the mean spectral density
\cite {Kesten,McKay}
\begin{eqnarray} \hspace{-10mm}
\label{eq:Mckay} \rho_{KM}(\mu) = \lim_{V\rightarrow\infty}
\langle\rho(\mu)\rangle=\left\{\begin{array}{lcr}\frac{d}{2\pi}
\frac{\sqrt{4(d-1)-\mu^2}}{d^2-\mu^2} & \mbox{for} &
|\mu|\le2\sqrt{d-1}\\ \\ 0 & \mbox{for} &
|\mu|>2\sqrt{d-1}\end{array}\right. \ .
 \end{eqnarray}
Note that the Kesten-McKay density depends explicitly on the degree
$d$. $\tilde{\rho}(\mu)$ is the fluctuating part of $\rho(\mu)$,
with $\left \langle\tilde{\rho}(\mu)\right \rangle = 0$. To simplify
the notation we omit reference to the $d$ dependence. In the limit
$d\gg1$, $\rho_{KM}$ approaches Wigner's semi-circle distribution
which characterizes the canonical Gaussian random matrix ensembles.
The fluctuating parts are expressed as infinite sums over Chebyshev
polynomials (of the first kind) $T_n(x)$  with coefficients which
depend on cycles on the graph. The trace formulae take similar forms
for the two ensembles, but with different coefficients.
\begin{eqnarray}\hspace{-10mm}
 \tilde{\rho}^{(A)}(\mu)
 &=&\
\frac{1}{\pi}\sum_{t=3}^\infty \frac{y^{(A)}_t}{\sqrt{4(d-1)-\mu^2}}
T_t \left( \frac{\mu}{2\sqrt{(d-1)}}\right).
   \label{eq:trace formula}
\end{eqnarray}
The parameters $y_t^{(A)}$ are defined as
\begin{equation}
y^{(A)}_t= \frac{1}{V} \frac{Y^{(A)}_t-(d-1)^t}{(\sqrt{d-1})^t}
  \label{eq:y_t}
\end{equation}
with $Y^{(A)}_t$ the number of $t$-periodic walks where no back
scattering is allowed (nb walks). Since $\left \langle
Y^{(A)}_t\right\rangle =(d-1)^t$,\  $y^{(A)}_t$ is the properly
regularized deviation of the number of $t$-periodic nb-walks from
their mean. In combinatorial graph theory it is customary to define
$C_t=\frac{Y^{(A)}_t}{2t}$ as the number of $t$-periodic nb cycles.
It is known that for $t<\log_{d-1}V$ and asymptotically in $V$, the
$C_t$ are distributed like independent Poisson variables
\cite{Janson,Wormald,Bollobas,McKayWormald}.

The trace formula for the spectral density of magnetic graphs is
similar to (\ref {eq:trace formula}) with the following differences:
The entire spectrum of the magnetic adjacency matrix is included in
the definition of the spectral density.  The coefficients
$y^{(M)}_t$ are now defined as
\begin{equation}
y^{(M)}_t = \frac{1}{V} \frac{ Y_t^{(M)}} {(\sqrt{(d-1)})^t}\ \ ,\ \
Y_t^{(M)}=\sum_{\alpha}{\rm e}^{i\chi_{\alpha}},
  \label{eq:y_t Magnetic}
\end{equation}
where the sum above is over all the nb t-periodic walks, and
$\chi_{\alpha}$ is the total phase (net magnetic flux) accumulated
along the t-periodic walk. For finite $V$, $\left \langle
y^{(M)}_t\right\rangle \ne 0$ since $\chi_{\alpha}=0$ for nb
periodic walks where each bond is traversed equal number of times in
the two directions. However, in the limit of large $V$, the number
of such walks is small and therefore $\left \langle
y^{(M)}_t\right\rangle \rightarrow 0$.

\subsection{Spectral fluctuations on graphs and RMT- Numerical
evidence}
 So far, the only evidence suggesting a connection between
RMT and the spectral statistics on graphs is the numerical studies
of Jacobson {\it et. al.} \cite{Jacobson}. In a preliminary step in
the present research, we performed numerical simulations which
extended the tests of \cite{Jacobson}. While describing these
studies, we shall introduce a few concepts from RMT which will be
used in the main body of the paper.

It is advantageous to map the spectrum from the real line to the
unit circle,
\begin{equation}
\phi_j =\arccos\frac{\mu_j}{2\sqrt{d-1}}\ \ ;\ \ 0\le \phi_j\le \pi\
.
\end {equation}
This change of variables is allowed since in  the limit of large
graphs, only a fraction of order $1/V$ of the spectrum is outside
the support of the Kesten McKay distribution
$[-2\sqrt{d-1},2\sqrt{d-1}]$ \cite{SashaSodin}.

The mean spectral density on the circle is not uniform, and the
Kesten McKay density on the circle is
\begin{equation}
\rho_{KM}(\phi) = \frac{2(d-1)}{\pi d}\frac{\sin^2\phi}
{1-\frac{4(d-1)}{d^2}\cos^2\phi}\ .
\end{equation}

The mean spectral counting function is defined as
\begin{equation}
\hspace{-17mm} N_{KM}(\phi)=V \int_0^{\phi }\rho_{KM}(\phi) d\phi\ =
V\frac{d}{2\pi}\left(\phi-\frac{ d-2 }{d}\arctan\left
(\frac{d}{d-2}\tan\phi\right )\right). \label{eq:counting}
\end{equation}

 Following the standard methods of spectral statistics, one
introduces a new variable $\theta$, which is uniformly distributed
on the unit circle. This ``unfolding" procedure is explicitly given
by
\begin{equation}
\theta_j= \frac{2\pi}{V}N_{KM}(\phi_j)
  \label{eq: definition theta_j}
\end{equation}

\begin{figure}[h]
  \centering
  \scalebox{0.8}{\includegraphics{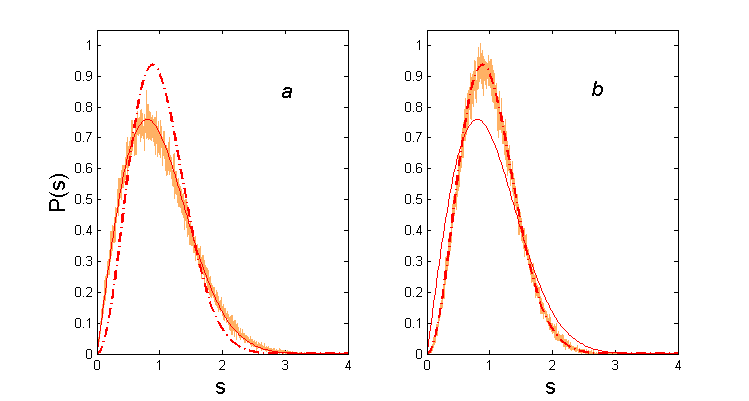}}
 \caption{Nearest level spacings for: (a.) Graphs possessing time reversal symmetry. (b.) Magnetic graphs.\\
 Both figures are accompanied with the RMT predictions: Solid line - COE, Dashed line - CUE.}
 \label{fig:Nearest Level Spacings}
\end{figure}

The nearest spacing distribution defined as
\begin{equation}
P(s) = \lim_{V\rightarrow\infty}\ \frac{1}{V} \left \langle
\sum_{j=1}^{V}\delta \left
(s-\frac{V}{2\pi}(\theta_j-\theta_{j-1})\right)\right\rangle\ ,
\end{equation}
is often used to test the agreement with the predictions of RMT
(This was also the test conducted in (\cite{Jacobson})). In this
definition of the nearest spacing distribution, $\theta_0$ coincides
with $\theta_V$, since the phases lie on the unit circle. In figure
(\ref{fig:Nearest Level Spacings}) we show numerical simulations
obtained by averaging over 1000 randomly generated $3$-regular
graphs on $1000$ vertices and their ``magnetic" counterparts,
together with the predictions of RMT for the COE and the CUE
ensembles \cite {Haakebook}, respectively. The agreement is quite
impressive.

Another quantity which is often used for the same purpose is the
spectral form-factor,
\begin{equation}
K_{V}(t)=\frac{1}{V}\left\langle \left|\sum_{j=1}^{V}{\rm
e}^{it\theta_j}\ \right|^2\ \right\rangle \ .
  \label{eq:form factor definition}
\end{equation}
The form-factor is the Fourier transform of the spectral two point
correlation function and it plays a very important r\^ole in the
understanding of the relation between RMT and the quantum spectra of
classically chaotic systems \cite{Haakebook, BerryA400}.

In RMT the form factor displays scaling: $\displaystyle
\lim_{V\rightarrow \infty}K_V(t)=K(\tau \equiv \frac{t}{V})$. The
explicit limiting expressions for the COE and CUE ensembles are
\cite {Haakebook}:
\begin{eqnarray} \hspace{-10mm}
\label{eq:Mckay} K_{COE}(\tau) =
\left\{\begin{array}{lcr}2\tau-\tau\log{(2\tau+1),} & \mbox{for} & \tau<1\\
\\ 2-\tau\log{\frac{2\tau+1}{2\tau-1},} & \mbox{for} & \tau>1\end{array}\right. \ .
  \label{eq: form factor COE}
 \end{eqnarray}

\begin{eqnarray} \hspace{-10mm}
\label{eq:Mckay} K_{CUE}(\tau) = \left\{\begin{array}{lcr}\tau, &
\mbox{for} & \tau<1\\ \\ 1, & \mbox{for} & \tau>1\end{array}\right.
\ . \label{eq: form factor CUE}
 \end{eqnarray}

\begin{figure}[h]
  \centering
  \scalebox{0.8}{\includegraphics{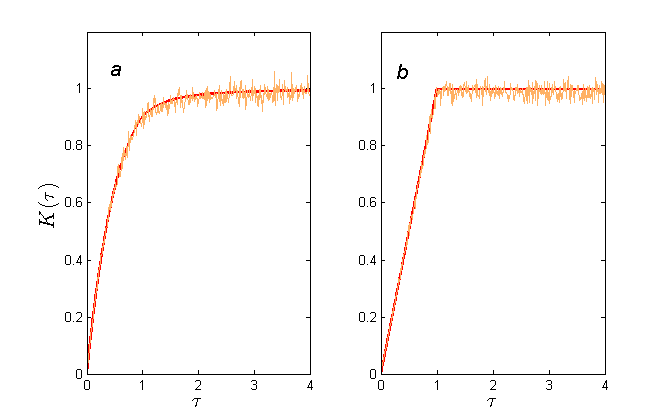}}
 \caption{The form factor $K(\tau)$ (unfolded spectrum) for (a.) $3$-regular graphs numerical \emph{vs.} the COE prediction.
 (b.) $3$-regular magnetic graphs \emph{vs.} the  CUE prediction.}
 \label{fig:Form Factors}
\end{figure}

The numerical data used to compute the nearest neighbor spacing
distribution $P(s)$, was used to calculate the corresponding form
factors for the non-magnetic and the magnetic graphs, as shown in
figure (\ref{fig:Form Factors}). The agreement between the numerical
results and the RMT predictions is apparent. This numerical data
triggered the research which is reported in the present article.

The above comparisons between the predictions of RMT and the
spectral statistics of the eigenvalues of $d$-regular graphs was
based on the unfolding of the phases $\phi_j$ into the uniformly
distributed phases $\theta_j$. As will become clear in the next
sections, it is more natural to study here the fluctuations in the
original spectrum and in particular the form factor
\begin{equation}
\widetilde{K}_{V}(t)=\frac{1}{V}\left\langle
\left|\sum_{j=1}^{V}{\rm e}^{it\phi_j}\ \right|^2\ \right\rangle \ .
  \label{eq:raw form factor definition}
  \end{equation}
The transformation between the two spectra was effected by (\ref{eq:
definition theta_j}) which is one-to-one and its inverse is defined:
\begin{equation}
\phi=S(\theta) \doteq N_{KM}^{-1}\left (V\frac{\theta}{2\pi}\right
)\ .
 \label{eq:mapping}
\end{equation}
This relationship enables us to express  $\widetilde{K}_{V}(t)$ in
terms of $\ K _{V}(t)$. In particular, if $K_{V}(t)$ scales by
introducing $\tau = \frac{t}{V}$ then,
\begin{equation}
  \label{eq:K_tilde_and_K}
\widetilde{K}(\tau=\frac{t}{V}) = \frac{1}{\pi}\int_0^{\pi} d\theta
K\left (\tau S^{'}(\theta)\right).
\end{equation}
The derivation of this identity is straightforward, and is described
in  \ref {appendix1}.

\begin{figure}[h]
  \centering
  \scalebox{0.8}{\includegraphics{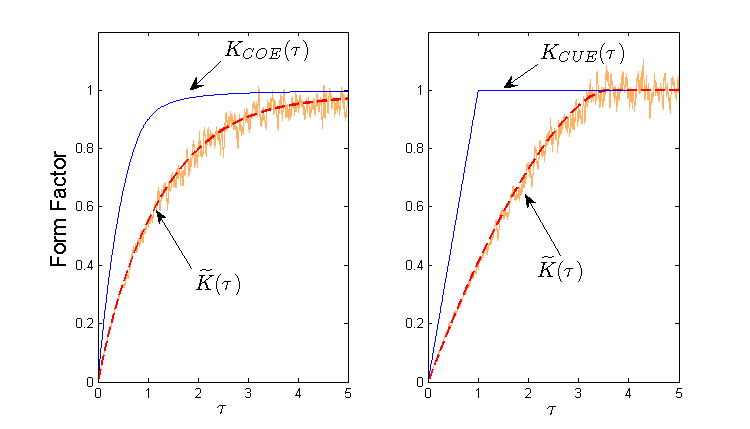}}
 \caption{The form factor $\widetilde{K}(\tau)$ (original spectrum) for 10-regular graphs. The numerical results are presented \emph{vs.}
 the expression (\ref{eq:K_tilde_and_K}) assuming RMT in the dashed line, and $K(\tau)$ (\ref{eq: form factor COE}, \ref{eq: form factor CUE}) in the solid line.}
 \label{fig:original Form Factors}
\end{figure}

Figure (\ref{fig:original Form Factors}) shows $\widetilde
{K}(\tau=\frac{t}{V})= \widetilde{K}_{V}(t)$ computed by assuming
that its unfolded analogue takes the RMT form (\ref {eq: form factor
COE}) or (\ref {eq: form factor CUE}), and it is compared with the
numerical data for graphs with $d = 10$. It is not a surprise that
this way of comparing between the predictions of RMT and the data,
shows the same agreement as the one observed previously.

\noindent \emph{Remark:} The rather unusual definition of the form
factor for the original spectrum can be illustrated by the following
example. Consider the Gaussian ensemble of $N$ dimensional symmetric
matrices (GOE). Its spectrum (properly normalized) is supported on
the interval $[-1,1]$ and the mean spectral density is given by
Wigner's semi-circle law. Mapping the spectrum onto the unit circle
results in points which are non-uniformly distributed.  One can
generate the form factors $K(\tau)$ and $\widetilde{K}(\tau)$ from
the original and unfolded spectra, and compare the numerical
distributions to the predictions from COE. The corresponding
$\widetilde{K}(\tau)$ is obtained from (\ref{eq:K_tilde_and_K}) in
the limit $d\gg 1$.

With this summary of definitions and numerical data we prepared the
background for the main results of the present work, where we use
the trace formulae to express the spectral form factor in terms of
the variance of the fluctuations in the counting of t-periodic nb
walks on the two graph ensembles. In Chapter \ref{sec:diagonal} we
shall use the known properties of $t$-periodic nb walks to compute
the leading term in the Taylor series of $\widetilde{K}(\tau)$, near
$\tau=0$. In Chapter \ref{sec:counting statistics} we shall take the
opposite direction, and by assuming that the spectral fluctuations
for the graphs are given by RMT, we shall derive new expressions for
the counting statistics of t-periodic nb walks on graphs. This
approach is similar in spirit to the work of Keating and Snaith
\cite{KeatingSnaith} who computed the mean moments of the Riemann
$\zeta$ function on the critical line, assuming that the
fluctuations of the Riemann zeros follow the predictions of RMT for
the CUE ensemble.

\section{From counting statistics of $t$-periodic orbits to RMT}
  \label{sec:diagonal}
In this section we shall establish a rigorous connection between the
spectral properties of regular graphs and those predicted by RMT. To
achieve this goal, we  use the trace formula (\ref{eq:trace
formula}), where $y_t$ are defined by (\ref{eq:y_t}) or (\ref{eq:y_t
Magnetic}) for the two ensembles.
\begin{equation}
\tilde{\rho}(\mu) = \frac{1}{\pi}\sum_{t=3}^\infty
\frac{y_t}{\sqrt{4(d-1)-\mu^2}} T_t\left( \frac{\mu}{2\sqrt{d-1}}
\right)  \nonumber
\end{equation}
Defining $\phi \equiv \arccos\left(\frac{\mu}{2\sqrt{d-1}}\right)$,
and using the orthogonality of the cosine, we can extract $y_t$,
\begin{equation}
y_t =  2\int_0^{\pi} \cos{(t \phi)} \tilde{\rho}(\phi) d\phi
\end{equation}
And so:
\begin{equation}
  \label{eq:(y_t)^2}
 \left \langle y^{2}_t \right \rangle =
 4\int_0^{\pi} \int_0^{\pi} \cos{(t \phi)} \cos{(t \psi)} \left \langle \tilde{\rho}(\phi)
\tilde{\rho}(\psi)\right \rangle d\phi d\psi \ .
\end{equation}
Recalling (\ref{eq:K_tilde_normalized _to_V})
\begin{equation}
  \label{eq:K_tilde_normalized _to_unity}
\widetilde{K}_{V}(s) \equiv 2V \int_0^{\pi} \int_0^{\pi} \cos(s
\phi) \cos(s \psi) \langle \tilde{\rho}(\phi) \tilde{\rho}(\psi)
\rangle d\phi d\psi\ ,
\end{equation}
and comparing (\ref{eq:(y_t)^2}) and (\ref{eq:K_tilde_normalized
_to_unity}) we get:
\begin{equation}
\widetilde{K}_{V}(t) = \frac{V}{2}\left \langle y^{2}_{t} \right
\rangle\ .
  \label{eq:K_tilde and y_t}
\end{equation}
So far the treatment of the two ensembles was carried on the same
formal footing. We shall now address each ensemble separately.

\subsection{The form factor for the $\mathcal{G}_{V,d}$ ensemble}
As was mentioned previously, it is useful to define the number of nb
t-cycles on the graph as $C_t = \frac{Y^{(A)}_t}{2t}$ (in this
definition, one does not distinguish between cycles which are
conjugate to each other by time reversal).  From combinatorial graph
theory it is known that on average $\left\langle C_t\right \rangle =
\frac{(d-1)^t}{2t}$ \cite{Janson,Wormald,Bollobas,McKayWormald}.
Hence $\widetilde{K}$ can also be written as:
\begin{equation}
\widetilde{K}^{(A)}_{V}(t) =
\frac{t}{V}\cdot\frac{\left\langle\left(C_{t}-\left\langle C_{t}
\right\rangle\right)^2\right\rangle}{\left\langle C_{t}
\right\rangle } \ .
  \label{eq:K tilde and var over mean}
\end{equation}
The expression of the spectral form factor
$\widetilde{K}^{(A)}_{V}(t)$ in terms of combinatorial quantities is
the main result of the present work. In particular, it shows that
the form factor is the ratio between the variance of $C_{t}$-the
number of nb t-cycles - and its mean. This relation is valid for all
$t$ in the limit $V\rightarrow \infty$.

For $t$ satisfying  $t<log_{d-1}V$, it is known that asymptotically,
for large $V$,  the $C_t$'s are distributed as independent Poisson
variables. For a Poisson variable, the variance and mean are equal.
This implies that for $\tau=\frac{t}{V}\rightarrow 0$,
$\widetilde{K}^{(A)}_{V}(t) = \tau$. Notice that the relation
(\ref{eq:K_tilde_and_K}) implies $K_{V}(t) = 2\widetilde{K}_{V}(t)$
for $\tau \rightarrow 0$ (see also \ref{eq: derivative form factor
at zero}).

Thus, for $\tau\ll1$
\begin{equation}
K^{(A)}(\tau) = 2 \tau \ .
\end{equation}

This result coincides with the COE prediction (\ref{eq: form factor
COE}). It provides the first rigorous support of the connection,
established so far numerically, between RMT and the spectral
statistics of graphs . It is analogous to Berry's ``diagonal
approximation" \cite{BerryA400} in quantum chaos.

\subsection{The form factor for the $\mathcal{G}^M_{V,d}$ ensemble}

In the magnetic ensemble the matrices (\ref{magneticA}) are complex
valued and Hermitian, which is tantamount to breaking time reversal
symmetry. The relevant RMT ensemble in this case is the CUE.

In the ensuing derivation we shall take advantage of the statistical
independence assumed for the magnetic phases which are uniformly
distributed on the circle. Ensemble averaging  will imply averaging
over both the magnetic phases and the graphs.

Recall that $y^{(M)}_t$, in the case of magnetic graphs, was
defined, by (\ref{eq:y_t Magnetic}):
\begin{eqnarray}
\label{eq:remind yt}
 y^{(M)}_t = \frac{1}{V} \frac{ Y_t^{(M)}}
{(\sqrt{(d-1)})^t}\ \ ,\ \ Y^{(M)}_t =\sum_{\alpha} {\rm
e}^{i\chi_{\alpha}}. \nonumber
\end{eqnarray}
$Y^{(M)}_t$ is the sum of interfering phase factors contributed by
the individual nb $t$-periodic walks on the graph. The phase factors
of periodic walks which are related by time reversal are complex
conjugated. Periodic walks which are self tracing (meaning that
every bond on the cycle is traversed the same number of times in
both directions), have no phase: $\chi_{\alpha}=0$. Using standard
arguments from combinatorial graph theory one can show that for $t<
\log_{d-1}V$, self tracing nb t-periodic walks are rare. Moreover,
the number of $t$-periodic walks which are repetitions of shorter
periodic walks can also be neglected. Hence
\begin{equation}
Y_t^{(M)} \approx 2 t \textstyle
\sum_{\alpha}^{'}\cos(\chi_{\alpha})
  \label{eq:trY^t magnetic}
\end{equation}
where $\sum^{'}$ includes summation over the nb $t$-cycles excluding
self tracing and non-primitive cycles. The number of $t$-cycles on
the graph is $C_t$, hence (\ref{eq:trY^t magnetic}) has
approximately $C_t$ terms. From (\ref{eq:trY^t magnetic}) and the
definition of $y_t^{(M)}$, it is easily seen that $\langle
(y^{(M)}_{t})^2 \rangle = \frac{1}{V^2 (d-1)^{t}}4t^2 \langle
 \left ( \sum^{'}\cos(\chi_{\alpha}) \right )^2 \rangle$. Averaging over the independent
magnetic phases we get that
 \begin{equation}
\widetilde{K}^{(M)}_{V}(t) = \frac{V}{2}\langle (y^{(M)}_{t})^2
\rangle \approx \frac{t}{2V} \equiv \frac{\tau}{2}.
  \label{eq: K_tilde magnetic DA1}
\end{equation}
For $\tau \rightarrow 0$, and to leading order, $K = 2\widetilde{K}$
(\ref{eq: derivative form factor at zero}). Hence
\begin{equation}
K^{(M)}(\tau) = \tau \ ,
  \label{eq: K_tilde magnetic DA2}
\end{equation}
which agrees with the CUE prediction.

\section{Counting statistics of $t$-periodic cycles on $d$-regular graphs from RMT}
  \label{sec:counting statistics}

In the previous section we made use of the known asymptotic
statistics of $Y_t$ to show that the leading term in the expansion
of $K_{V}(t)$ behaves as $g\tau$ where $g=1,2$ for the two graph
ensembles. This property is consistent with the predictions of RMT.
Had we known more about the counting statistics, we could make
further predictions and compare them to RMT results. However, to the
best of our knowledge we have exhausted what is known from
combinatorial graph theory, and the only way to proceed would be to
take the reverse approach, and \emph{assume} that the form factor
for graphs is given by the predictions of RMT, and see what this
implies for the counting statistics. Checking these predictions from
the combinatorial point of view is beyond our scope. However, we
shall show that they are accurately supported by the numerical
simulations.

The starting points for the discussion are the relations
(\ref{eq:K_tilde and y_t}) and (\ref{eq:K_tilde_and_K}) which can be
combined to give
\begin{equation}
 \left\langle (y_{t})^2\right\rangle = \frac{2}{V}\widetilde{K}_{V}(t)
 =\frac{2}{V\pi}\int_0^{\pi}
d\theta K\left (\tau S^{'}(\theta)\right)\ .
  \label{eq:y_t and K}
\end{equation}
Our strategy here will be to use for the unfolded form factor the
known expressions from RMT (\ref{eq: form factor COE}) and (\ref{eq:
form factor CUE}) and compute $\left\langle (y_{t})^2\right\rangle
$. This will provide an expression for the combinatorial quantities
defined for each of the graph ensembles, and expanding in $\tau$ we
shall compute the leading correction to their known asymptotic
values.

To proceed, we have to analyze the integral (\ref{eq:K_tilde_and_K})
and expand it in powers of  $\tau$ near $\tau=0$. For this purpose
we have to recall the function $S(\theta) =N_{KM}^{-1}\left
(V\frac{\theta}{2\pi}\right )$ (\ref{eq:mapping}). The inversion of
the spectral counting function needs more attention near the end
points of the support, where
\begin{equation}
  \label{eq:phi spectral counting function}
\hspace{-2cm} N_{KM}(\phi) \underset{\phi\rightarrow
0}\longrightarrow\frac{2V}{3\pi}D\phi^3;\ \ D \equiv
\frac{d(d-1)}{(d-2)^2}\ .
\end{equation}
Thus, in the vicinity of $\theta =0$
\begin{equation}
 S(\theta) \approx S_0(\theta) = \left[ \frac{3\theta}{4D}
\right]^{\frac{1}{3}},
\end{equation}
and $S'$ which is singular near $0$ is
\begin{equation}
  \label{eq:S^{'}}
S^{'}(\theta) \approx S_0^{'}(\theta) = \frac{1}{4D}\
\left[\frac{4D}{3\theta}\right]^{\frac{2}{3}}.
\end{equation}
Both RMT form factors take the value $1$ for argument values
sufficiently larger than $1$. Hence the value of $\theta$ where the
$\tau S'(\theta) =1$ plays an important r\^ole. We shall denote it
by $\theta_m(\tau)$, and for small values of $\tau$ it takes the
value
 $$\theta_m(\tau) =
\frac{4D}{3}\cdot\left[\frac{\tau}{4D}\right]^{\frac{3}{2}}$$.

This information suffices for the analytic derivations which will
follow. However, for the numerical computation of
$\widetilde{K}(\tau)$ for the entire range of $\tau$, we shall need
a better approximation for $S(\theta)$ valid over the entire range
of integration. This can be achieved by successive Newton-Raphson
iterations. To second order,
\begin{equation}
\hspace{-26mm} S(\theta) = S_0+\frac{1}{8d(d-1)}\left(2\theta-2d
S_0+2(d-2)\arctan   \frac{d\tan \left(S_0\right)}{d-2} \right )\cdot
\left( d^2+(d-2)^2\cot^2 \left(S_0\right) \right).
  \label{eq:unfolding phi into theta}
\end{equation}
The numerical error, induced by this approximation, is less than 5
percents, (it is less than 2 percents for $d>8$).

We now turn to the small $\tau$ domain. We shall treat the two
ensembles separately starting with the magnetic ensemble since it is
simpler.

\subsection{Counting statistics for the $\mathcal{G}^M_{V,d}$ ensemble}
The CUE form factor (\ref{eq: form factor CUE}) is: $K_{CUE}(\tau) =
\tau$ for $\tau<1$, and $K_{CUE}(\tau) = 1$ for $1 \le \tau$.
Therefore, we can divide the integral (\ref{eq:K_tilde_and_K}), in
the following way:
\begin{eqnarray}
\widetilde{K}^{(M)}(\tau) = \frac{1}{\pi}\left(\int_0^{\theta_m}
d\theta+ \int_{\theta_m}^\pi \tau S^{'}(\theta)d\theta \right) =
\frac{1}{\pi} \left(\theta_m+\frac{\pi\tau}{2}-\tau
S(\theta_m) \right) = \\
= \frac{\tau}{2}+\frac{1}{\pi} \left(\theta_m-\tau S(\theta_m)
\right) \nonumber
\end{eqnarray}
Hence, the first two terms in the expansion of $\widetilde{K}(\tau)$
are
\begin{equation}
  \label{eq:K tilde GUE expansion}
\widetilde{K}^{(M)}(\tau) =
\frac{\tau}{2}+f_1(d)\tau^{\frac{3}{2}}+\ldots
\end{equation}
where
\begin{equation}
f_1(d) = -\frac{1}{3\pi\sqrt{D}}\ .
  \label{eq:f_1 GUE}
\end{equation}
Thus, the difference
$(\widetilde{K}^{(M)}(\tau)-\frac{\tau}{2})/f_1(d)$, should scale
for small $\tau$  as $\tau^{\frac{3}{2}}$ for all values of $d$.
This data collapse is shown in  figure (\ref{fig:Data collapse
GUE}).
\begin{figure}[h]
  \centering
  \scalebox{0.8}{\includegraphics{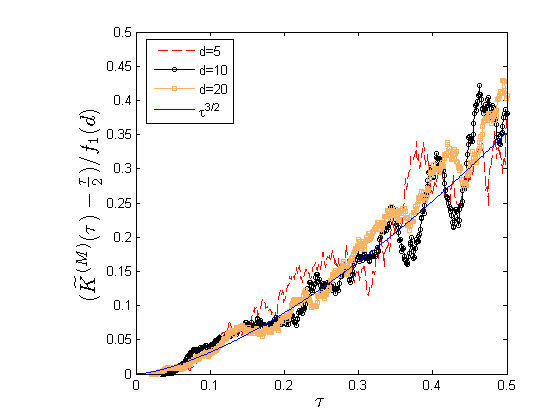}}
 \caption{$(\widetilde{K}^{(M)}(\tau)-\frac{\tau}{2})/f_1(d)$ for various values of $d$ \emph{vs.} the curve $\tau^{\frac{3}{2}}$.}
 \label{fig:Data collapse GUE}
\end{figure}

\subsection{Counting statistics for the $\mathcal{G}_{V,d} $ ensemble}
Similar results can be obtained for the counting statistics of the
$\mathcal{G}_{V,d} $ ensemble. Here, the relevant RMT form factor is
the COE expression (\ref{eq: form factor COE}). The  integral
(\ref{eq:K_tilde_and_K}), is divided in the following way:
\begin{eqnarray}
\widetilde{K}^{A}(\tau) = \frac{1}{\pi}\left(\int_0^{\theta_m}
(2-\tau S^{'}(\theta)\log{\frac{2\tau S^{'}(\theta)+1}{2\tau
S^{'}(\theta)-1}}) d\theta\right)\\  + \frac{1}{\pi} \left(
\int_{\theta_m}^\pi 2\tau S^{'}(\theta)-\tau
S^{'}(\theta)\log{(2\tau S^{'}(\theta)+1)}
d\theta\right) \\
= \frac{1}{\pi}\left(2\theta_m+\pi\tau-2\tau S(\theta_m)-\tau
\int_0^{\theta_m} S^{'}(\theta)\log{\frac{2\tau
S^{'}(\theta)+1}{2\tau S^{'}(\theta)-1}} d\theta \right)\\
+\frac{1}{\pi}\left(-\tau \int_{\theta_m}^\pi
S^{'}(\theta)\log{(2\tau S^{'}(\theta)+1)} d\theta\right),
\end{eqnarray}
We change variables to integrate over the new variable $x \equiv
S(\theta)$. Denoting $S^{'}(\theta)$ by $f(x)$, We get:
\begin{eqnarray}
  \label{eq:K tilde GOE after change of variables}
\widetilde{K}^{(A)}(\tau) =
\frac{1}{\pi}\left(2\theta_m+\pi\tau-2\tau S(\theta_m)-\tau
\int_0^{S(\theta_m)} dx\log{\frac{2\tau f(x)+1}{2\tau f(x)-1}}
\right)\\ \nonumber +\frac{1}{\pi}\left(-\tau
\int_{S(\theta_m)}^{\frac{\pi}{2}} dx\log{(2\tau f(x)+1)} \right),
\end{eqnarray}
At small values of $\tau$ (which imply small values of $\theta_m$),
$f(x)=\frac{1}{4Dx^2}$. The first integral in (\ref{eq:K tilde GOE
after change of variables}) can be solved explicitly. In the second
integral, we can restrict ourselves only to the interval in which
$f(x) \propto\frac{1}{x^2}$ because the rest of the integral will
only yield higher order terms in $\tau$. Therefore, we can also
solve the second integral. Finally we get
\begin{equation}
  \label{eq:K tilde GOE expansion}
\widetilde{K}^{(A)}(\tau) = \tau+f_2(d)\tau^{\frac{3}{2}}+\ldots
\end{equation}
where
\begin{equation}
f_2(d) = \frac{1}{\sqrt{2D}} \left(
\frac{2}{\pi}\cdot\text{arccoth}(\sqrt{2})-\frac{2\sqrt{2}}{3\pi}-1
\right).
  \label{eq:f_2 GOE}
\end{equation}
Thus, the difference $(\widetilde{K}^{(A)}(\tau)-\tau)/f_2(d)$, at
small $\tau$, should scale as $\tau^{\frac{3}{2}}$ independently of
$d$. This data collapse is shown in figure (\ref{fig:Data collapse
GOE}).
\begin{figure}[h]
  \centering
  \scalebox{0.8}{\includegraphics{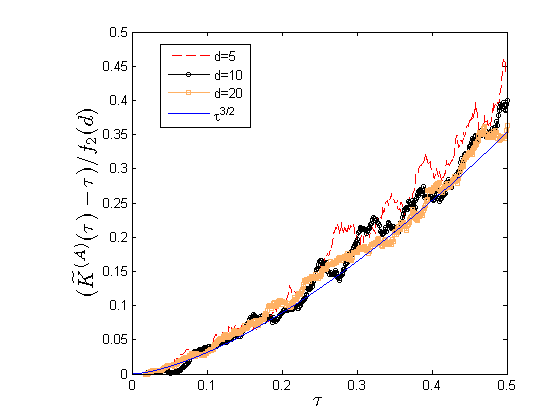}}
 \caption{$(\widetilde{K}^{(A)}(\tau)-\tau)/f_2(d)$ for various values of $d$ \emph{vs.} the curve $\tau^{\frac{3}{2}}$.}
 \label{fig:Data collapse GOE}
\end{figure}

Using the above and (\ref{eq:K tilde and var over mean}) we can
write,
\begin{equation}
  \hspace{-2cm}
\frac{\left\langle\left(C_{t}-\left\langle C_{t}
\right\rangle\right)^2\right\rangle}{\left\langle C_{t}
\right\rangle}
 = \frac{1}{\tau\pi}\int_0^{\pi} d\theta K_{COE}\left (\tau
S^{'}(\theta)\right)\underset{\tau \rightarrow
0}\longrightarrow1+f_2(d)\sqrt{\tau}+\ldots
\end{equation}
If the $C_t$'s were Poissonian random variables, the expansion above
would terminate at $1$. Since it does not, we must conclude that the
$C_t$'s are not Poissonian. The highest order deviation comes from
the next order term in the expansion which is proportional to
$\tau^{\frac{1}{2}}$. The coefficient, $f_2(d)$, is explicitly
calculated above.

We can examine the behavior at another domain of $\tau$, namely
$\tau>1$.\\
It can easily be shown that $S^{'}(\theta)\geq \frac{d}{4(d-1)}$.
Consequently, for $\tau>\frac{4(d-1)}{d}$, the argument of $K$ in
(\ref{eq:K_tilde_and_K}) is larger than one, and so
\begin{equation}
\lim_{\tau\rightarrow \infty} \widetilde{K}^{(A)}(\tau) = 1.
\end{equation}
Combining this result with (\ref{eq:K tilde and var over mean}),
provides the asymptotic of the variance-to-mean ratio:
\begin{equation}
\lim_{\tau \rightarrow \infty}\tau \frac{var(C_{t})}{\langle
C_{t}\rangle}=1,\ \ \ $for$\  V,t \rightarrow \infty;\ \
\frac{t}{V}=\tau
\end{equation}
This is a new interesting combinatorial result, since very little is
known about the counting statistics of periodic orbits in the regime
of $\tau>1$.

\section{Discussion}

The two main results of the present paper can be summarized as
follows. First, we have shown that to leading order, the spectral
statistics for graphs with time reversal symmetry is consistent with
the COE, and when time reversal
symmetry is broken, the CUE statistics come to play.\\
Second, by inverting the argument, and assuming RMT for the spectral
statistics, we derived new results in graph theory, namely the
deviation of the number of cycles from complete randomness, and the
statistics at large $\tau$ . We do not know at this point how to
interpret these results from a combinatorial point of view. This
remains for now an open question. It is important to emphasize that
unlike the standard approach in RMT, in this paper we have worked
with the entire spectrum (bulk and edge states), not merely the
bulk. As a result, effects of the edge states must be taken into
account when trying to give a
combinatorial answer to the questions posed above.\\
So far our rigorous results are rather limited. Yet, this work paves
the way to further studies where the intricate relationship between
combinatorial graph theory and RMT will be elucidated.

\section*{Acknowledgments}

\noindent The authors wish to express their gratitude to Mr. Amit
Godel who was a co-author in the first paper in the series. We thank
him for many fruitful discussions, and for his careful reading of
the manuscript, which resulted in many fine remarks and
corrections.\\
The authors are also grateful to Mr. Sasha Sodin for many insightful
discussions and for his much needed assistance, throughout this
series
of papers.\\
Finally, the authors thank both referees, the first
referee in particular, for pointing out several errors.\\
This work was supported by the Minerva Center for non-linear
Physics, the Einstein (Minerva) Center at the Weizmann Institute and
the Wales Institute of Mathematical and Computational Sciences)
(WIMCS). Grants from EPSRC (grant EP/G021287), and BSF (grant
2006065) are acknowledged.

\appendix
\section {The relation between $K(\tau)$ and $ \widetilde{K} (\tau)$}
\label{appendix1}
 The form factor (\ref{eq:raw form factor definition}) can be rewritten in the form
\begin{eqnarray}
  \label{eq:K_tilde}
\widetilde{K}_{V}(s) &\equiv&  \frac{2}{ V} \left \langle \left |
\sum_{j=1}^{V} \cos(s \phi_j)\right |^2 \right \rangle
\nonumber \\
&=&
 2 V \int_0^{\pi} \int_0^{\pi} \cos(s \phi) \cos(s \psi)
\langle \rho(\phi) \rho(\psi) \rangle d\phi d\psi
\end{eqnarray}
The factor $V$ above is due to the normalization (\ref{eq:density})
of the spectral density. The smooth part of the spectral density
does not encode any information about spectral fluctuations, so we
are only interested in the fluctuating part, $\tilde{\rho}$. Thus,
\begin{equation}
  \label{eq:K_tilde_normalized _to_V}
\widetilde{K}_{V}(s) \equiv  2 V  \int_0^{\pi} \int_0^{\pi} \cos(s
\phi) \cos(s \psi) \langle \tilde{\rho}(\phi) \tilde{\rho}(\psi)
\rangle d\phi d\psi
\end{equation}
We emphasize again that the main difference between $\widetilde{K}$
and the actual form factor, comes from the fact that the $\phi$'s
are not uniformly distributed. Using the mapping $\phi=S(\theta)$
(\ref{eq:mapping}), we get:
\begin{equation}
  \hspace{-1cm}
\widetilde{K}_{V}(t) = 2V \int_0^{2\pi} \int_0^{2\pi}
\cos(tS(\theta)) \cos(tS(\theta^{'})) \langle \tilde{\rho}(\theta)
\tilde{\rho}(\theta^{'}) \rangle d\theta d\theta^{'} \nonumber
\end{equation}

The two-point correlation function is defined as $R_2(w)= \left( 2
\pi\right)^2 \left \langle
\tilde{\rho}(\theta)\tilde{\rho}(\theta+w\frac{2\pi}{V}) \right
\rangle$. In addition, we change variables to $\eta =
\frac{\theta+\theta^{'}}{2}, w = \theta-\theta^{'}$, and we expand
the integrand up to first order in $w$, keeping in mind that
$R_2(w)$ is only of significant magnitude if $w$ is small. We are
thus left with:
\begin{equation}
\hspace{-2cm} \widetilde{K}_{V}(t) =\frac{V}{4 \pi^2}\int_0^{2\pi}
d\eta \int_{-2\pi}^{2\pi}dw R_2\left (\frac{wV}{2 \pi}\right ) \cdot
\left[ \cos(2tS(\eta))+\cos(twS^{'}(\eta)) \right]
\end{equation}
The first integral is:
\begin{eqnarray}
\frac{V}{4 \pi^2}\int_0^{2\pi} d\eta \int_{-2\pi}^{2\pi}dw
R_2\left (\frac{wV}{2 \pi}\right )\cos(2tS(\eta))\\
= \frac{1}{2\pi}\int_0^{2\pi} d\eta \cos{(2tS(\eta))}
\int_{-\infty}^{\infty} ds R_2(s) = \delta(\tau)
\end{eqnarray}
where we used $\int_{-\infty}^{\infty} ds R_2(s) = 1$ (see for
example (\cite{GVZ89})).

The second integral is:
\begin{equation}
\hspace{-2cm} \frac{V}{4 \pi^2}\int_0^{2\pi} d\eta
\int_{-2\pi}^{2\pi}dw R_2\left (\frac{wV}{2 \pi}\right
)\cos(twS^{'}(\eta)) = \frac{1}{2\pi}\int_0^{2\pi} d\eta K\left
(\tau S^{'}(\eta)\right)
\end{equation}
And we conclude that:
\begin{equation}
\widetilde{K}_{V}(t) = \delta(\tau)+\frac{1}{2\pi}\int_0^{2\pi}
d\eta K\left (\tau S^{'}(\eta)\right)
  \label{relation K and K-tilde}
\end{equation}
Where $\tau$ is defined as before, and $\widetilde{K}_{V}(t)$ admits
the same scaling as in RMT: $\widetilde{K}(\tau=t/V)$. Finally, we
drop the $\delta$-function term, and we take advantage of the fact
that $S$ is symmetric around $\pi$ (this is a consequence of the
Kesten-McKay measure being symmetric around zero). This completes
the proof, and we end up with (\ref{eq:K_tilde_and_K}):
\begin{eqnarray}
\widetilde{K}_{V}(\tau) = \frac{1}{\pi}\int_0^{\pi} d\theta K\left
(\tau S^{'}(\theta)\right) \nonumber
\end{eqnarray}

Using this relation, we can prove that the slope of $K$ is twice
that of $\widetilde{K}$  at $\tau = 0$. Denote $\frac {{\rm
d}K(\tau=0)}{{\rm d}\tau} = g$. Then,
\begin{equation}
 \frac {{\rm d}\widetilde{K}(\tau=0)}{{\rm d}\tau}   =
\frac{g}{\pi}\int_0^\pi S^{'}(\theta)d\theta = \frac{g}{2}
  \label{eq: derivative form factor at zero}
\end{equation}
which proves the above.


\end{document}